\documentclass[sigconf,review]{acmart}

\usepackage{booktabs}
\usepackage{microtype}
\usepackage{balance}
\setcopyright{none}
\renewcommand\footnotetextcopyrightpermission[1]{}
\begin{document}

\title{eBandit: Kernel-Driven Reinforcement Learning for Adaptive Video Streaming}

\author{Mahdi Alizadeh}
\affiliation{%
  \institution{University of Southern California}
  \city{Los Angeles}
  \state{CA}
  \country{USA}
}
\email{malizade@usc.edu}

\begin{abstract}
User-space Adaptive Bitrate (ABR) algorithms cannot see the
transport layer signals that matter most, such as minimum RTT and
instantaneous delivery rate, and they respond to network changes only
after damage has already propagated to the playout buffer.
We present \emph{eBandit}, a framework that relocates both network
monitoring and ABR algorithm selection into the Linux kernel using eBPF.
A lightweight epsilon-greedy Multi-Armed Bandit (MAB) runs inside a
\texttt{sockops} program, evaluating three ABR heuristics against a reward derived from
live TCP metrics.
On an adversarial synthetic trace eBandit achieves
$416.3 \pm 4.9$ cumulative QoE, outperforming the best static
heuristic by $7.2\%$.
On 42 real-world sessions eBandit achieves a mean QoE per chunk of $1.241$, the highest
across all policies, demonstrating that
kernel-resident bandit learning transfers to heterogeneous mobile conditions.
\end{abstract}

\maketitle
\thispagestyle{empty}
\pagestyle{empty}

%%
%% 1. INTRODUCTION
%%
\section{Introduction}

The explosive growth of video-on-demand and live streaming has made
adaptive bitrate (ABR) selection one of the most performance-critical
decisions on the Internet.
Every time a user watches a video, a client-side algorithm must choose
the quality of the next segment based on its estimate of available
bandwidth.
Get it wrong, and the viewer suffers rebuffering stalls or unnecessary
quality drops.
Yet today's ABR algorithms make this decision entirely in user space,
often inside a browser sandbox that has no access to low-level network
signals. Meanwhile, the Linux kernel already tracks fine-grained TCP
state, such as smoothed RTT and per-ACK delivery rate, that would let
ABR react faster and more accurately.
Our core motivation follows naturally from this observation.
If the operating system already
knows the network state, the ABR decision should be made there.

Modern ABR clients execute entirely in user space, relying on
application-layer proxies such as throughput estimates and buffer level
as surrogates for the true network state.
This architecture suffers from two compounding problems.
First, an observability gap separates the kernel from user
space. By the time a throughput
drop or RTT spike propagates to the application layer, one or more video
segments have already been requested at an inappropriate bitrate and the
playout buffer has begun to drain.
The kernel's TCP control block maintains sub-millisecond RTT
measurements and per-ACK delivery-rate samples that are simply invisible
to user-space sockets, creating an order-of-magnitude timing gap
between the kernel's view and the application's.
Second, individual heuristics are brittle because no single static ABR
policy dominates across all network conditions.
Throughput-based schemes~\cite{yin2015mpc} over-react on jittery links;
buffer-based schemes~\cite{huang2014bba} stall catastrophically after
cliff-edge capacity drops; and utility-maximizing controllers such as
BOLA~\cite{spiteri2016bola} lag behind rapid regime shifts.
A recent measurement study by Yan et al.~\cite{yan2020puffer} confirms
this at scale, showing that in-situ conditions determine which
algorithm wins on any given session, making a static choice inherently
suboptimal.

We argue that \emph{policy selection} should happen where TCP state is
natively visible.
To this end we present \textbf{eBandit}, a framework built on the Linux
eBPF subsystem~\cite{vieira2020ebpf} that (i)~monitors TCP connections
at socket-event granularity via a \texttt{sockops} program,
(ii)~computes a per-RTT reward from raw kernel metrics using integer
arithmetic only, and (iii)~runs an epsilon-greedy Multi-Armed Bandit
(MAB) entirely in kernel space to select the ABR policy that performs
best in the current network regime.
The user-space client becomes a thin executor that polls a shared eBPF
map at chunk boundaries and dispatches to whichever policy the kernel
has elected, with zero recompilation and sub-microsecond overhead.

eBPF programs run in a sandboxed, JIT-compiled environment attached to
kernel hooks, providing \emph{zero-copy, zero-context-switch} access to
TCP control-block fields such as \texttt{srtt\_us}, delivery rate, and
bytes delivered, all of which are invisible to user-space sockets.
The epsilon-greedy MAB is deliberately chosen over deeper RL
models~\cite{abbasloo2020orca}: its state is three
$\langle\text{count},\,\text{total}\rangle$ pairs (one per arm) and
its update is a single integer addition, fitting within
eBPF stack and instruction limits.
A warm-start variant bootstraps from a small held-out trace set to skip
the cold-start exploration penalty in production.
Unlike learned ABR systems~\cite{mao2017pensieve} that require heavy
offline training and are difficult to update once deployed, eBandit
learns continuously from live traffic and can adapt to new network
regimes without retraining.
By combining the kernel-residence of recent eBPF
systems~\cite{zhou2023electrode,miano2023sketching} with online bandit
learning, eBandit occupies a design point between static heuristics
and heavyweight neural ABR.

%%
%% 2. SYSTEM DESIGN
%%
\section{System Design \& Approach}

\begin{figure}[t]
  \centering
  \includegraphics[width=\linewidth]{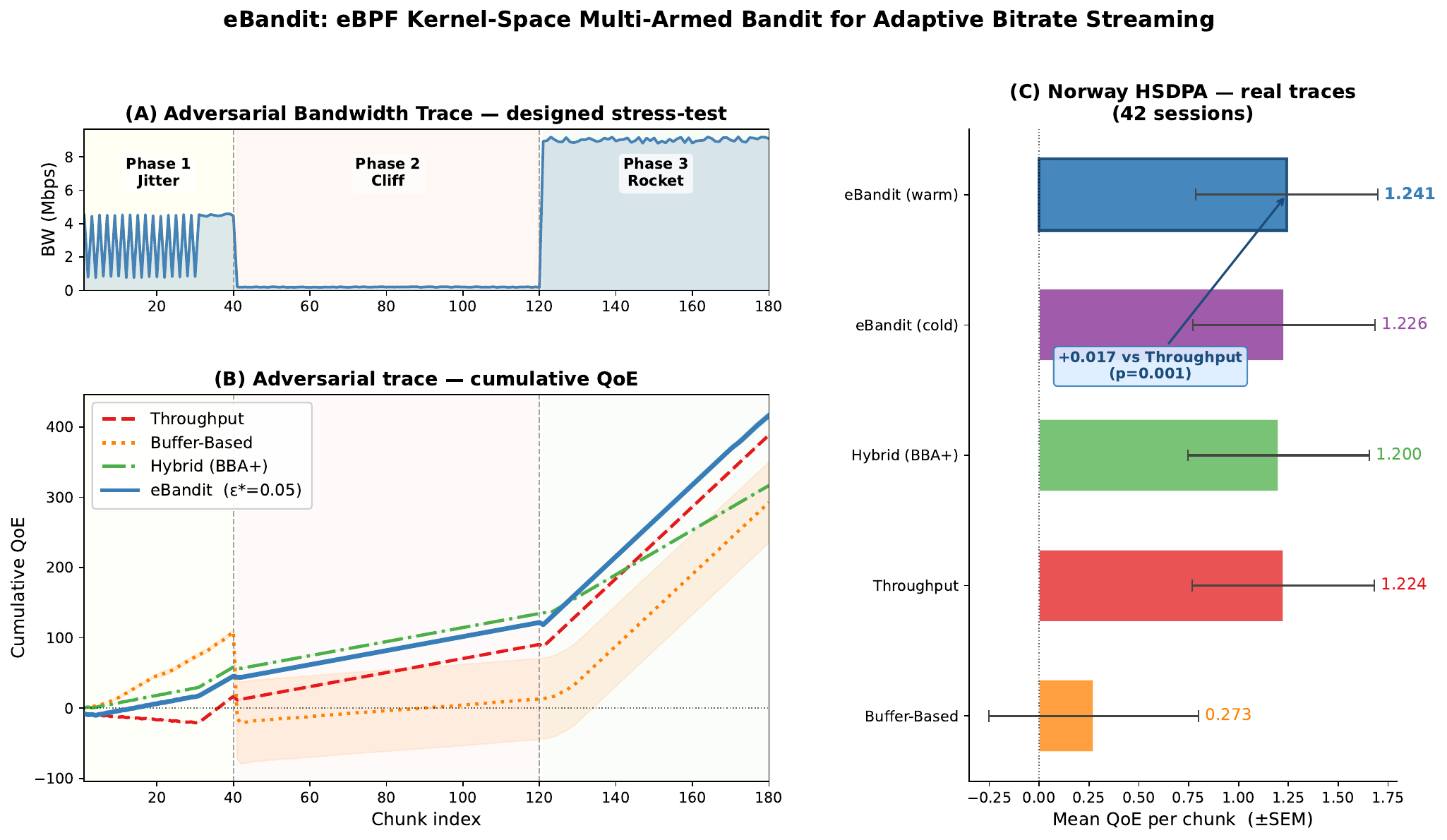}
  \caption{eBandit evaluation. \textbf{(A)}~Adversarial three-phase
    bandwidth trace (jitter $\to$ cliff $\to$ rocket).
    \textbf{(B)}~Cumulative QoE on the adversarial trace, averaged
    over 10 seeds; shaded bands are $\pm 1\sigma$.
    \textbf{(C)}~Mean QoE per chunk on 42 Norway HSDPA sessions
    ($\pm$SEM); eBandit (warm) achieves the highest score.}
  \label{fig:eval}
\end{figure}

\textbf{Kernel-Space Component.}
A \texttt{sockops} eBPF program is attached to the root cgroup,
hooking into \texttt{ACTIVE\_ESTAB\-LISHED\_CB} (to subscribe new
connections to RTT callbacks) and \texttt{RTT\_CB} (to receive
per-RTT measurements).
On every RTT callback the program reads three fields from the
\texttt{bpf\_sock\_ops} context, namely \texttt{srtt\_us},
\texttt{rate\_\allowbreak{}delivered}, and
\texttt{rate\_\allowbreak{}interval\_us}, and computes a normalized
reward using integer arithmetic only (the BPF verifier forbids
floating-point):
\begin{equation}
  r_t = \frac{w_r \cdot \frac{\mathit{rtt}_{\min}}{\mathit{rtt}_t}
        \;+\; w_d \cdot \frac{d_t}{d_{\max}}}{w_r + w_d},
  \label{eq:reward}
\end{equation}
where $\mathit{rtt}_{\min}$ is the running minimum RTT,
$d_{\max}$ is the running maximum delivery rate, and $w_r{=}6$,
$w_d{=}4$ weight latency (60\%) more heavily than throughput (40\%).
All values are scaled by a constant $S{=}1000$ to preserve precision
in fixed-point; the result $r_t \in [0,\,1000]$.
This reward is attributed to whichever ABR arm is currently active
(recorded in a per-IP \texttt{BPF\_MAP\_TYPE\_HASH} entry keyed by the
remote IPv4 address) and accumulated into the arm's
$\langle\text{pull\_count},\,\text{total\_reward}\rangle$ entry in a global
\texttt{BPF\_MAP\_TYPE\_ARRAY} of three arms.
The epsilon-greedy selection step executes using only 32-bit integer
arithmetic to respect eBPF's verifier constraints; stochastic arm
selection is seeded from \texttt{bpf\_get\_prandom\_u32()}.
Arm comparison avoids division entirely via cross-multiplication so that
$\bar{r}_i > \bar{r}_j$ iff
$\text{total}_i \times \text{count}_j > \text{total}_j \times
\text{count}_i$, and the loop over three arms is manually unrolled
because the BPF verifier rejects backward-edge loops in BCC-compiled
programs.

A \emph{shock detection} heuristic augments the MAB: when the
2-chunk rolling bandwidth average increases by $>2.5\times$,
the agent overrides exploration and selects the Throughput arm for a
cooldown window.
This mechanism is currently in user-space simulation; kernel-side
porting is planned.

\textbf{User-space component.}
The video client implements three independent policy modules:
Throughput (harmonic-mean estimator, window~3,
following~\cite{yin2015mpc}), BOLA (buffer-aware utility
maximization~\cite{spiteri2016bola} with adaptive safety margin), and
Hybrid BBA+ (buffer thresholds gated by a bandwidth cap,
extending~\cite{huang2014bba}).
At each chunk boundary the client issues a single
\texttt{bpf\_map\_lookup\_elem} call (sub-microsecond) to read the
elected arm ID and dispatches to the corresponding module.
Policy switching is stateless and no ABR internal state is carried across
arm transitions.
A warm-start path pre-populates the MAB's reward accumulators from a
small held-out trace set, eliminating the cold-start exploration
phase in latency-sensitive deployments.
Because warm-start only adjusts initial counts in the bandit's map
entries, it adds no runtime overhead.

\textbf{Overhead.}
The dominant cost is the per-RTT eBPF callback, which executes entirely
in kernel space without a context switch.
Based on published BPF JIT
characterisation~\cite{miano2023sketching,zhou2023electrode}, we
\emph{estimate} the full callback (reward computation + arm update) at
$\approx$300\,ns, which is two orders of magnitude below a user-space
\texttt{/proc/net/tcp} poll ($\approx$10--50\,µs).
The per-chunk \texttt{bpf\_map\_lookup\_elem} adds $<$1\,µs,
negligible relative to a 2.5\,s chunk duration.

%%
%% 3. PRELIMINARY EVALUATION
%%
\section{Evaluation}
\label{sec:eval}

We evaluate eBandit against three static baselines (Throughput-Based,
Buffer-Based (BBA)~\cite{huang2014bba}, and Hybrid BBA+) using a
purpose-built adversarial synthetic trace and the standard Norway HSDPA
mobile dataset~\cite{riiser2013} (Figure~\ref{fig:eval}).
All evaluations use a faithful Python simulation of the kernel MAB
logic (identical reward function, epsilon-greedy policy, and
arm-selection rules), enabling reproducible experiments over diverse
trace corpora without requiring root access.
We adopt the log-utility QoE model of
Pensieve~\cite{mao2017pensieve}:
\begin{equation}
  \mathrm{QoE} = \log_2\!\left(\frac{b_t}{150}\right)
    - \lambda \cdot \text{rebuf}
    - \mu \cdot |\Delta \log_2 b|,
  \label{eq:qoe}
\end{equation}
with $\lambda{=}4.3$ and $\mu{=}2.0$.

\textbf{Adversarial synthetic trace.}
eBandit outperforms the best static heuristic by
7.2\% and avoids the catastrophic stalls of buffer-based schemes.
The trace (Figure~\ref{fig:eval}A) stresses each heuristic's known
failure mode in three phases.
In Phase~1 (Jitter, chunks 1--40), bandwidth oscillates between
4.5 and 0.8\,Mbps, making throughput estimates volatile.
In Phase~2 (Cliff, chunks 41--120), capacity drops to 200\,kbps;
buffer-based schemes, having filled the buffer, select high-bitrate
segments and stall catastrophically~\cite{huang2014bba}.
In Phase~3 (Rocket, chunks 121--180), capacity surges to 9\,Mbps;
eBandit's shock detector promotes the Throughput arm for aggressive
quality ramp-up.
eBandit ($\varepsilon^{*}{=}0.05$) achieves cumulative
QoE of $416.3 \pm 4.9$, versus $388.5$ for Throughput, $316.7$ for
Hybrid, and $292.6 \pm 57.4$ for Buffer-Based, which is $7.2\%$ above
the best static policy (Figure~\ref{fig:eval}B).

\textbf{Real-world traces.}
On mobile traces, eBandit (warm) achieves the highest mean QoE
per chunk among all evaluated policies.
We evaluate on 42 sessions drawn from bus, tram, train, and ferry
subsets of the Norway HSDPA dataset~\cite{riiser2013}, withholding
8 traces for warm-start calibration.
Epsilon is fixed at $0.10$ (the Pensieve
default~\cite{mao2017pensieve}) and not tuned on HSDPA data to
avoid circularity.
eBandit (warm) achieves a mean QoE per chunk of $1.241$, compared with
$1.226$ for cold-start, $1.224$ for Throughput, $1.200$ for Hybrid,
and $0.273$ for Buffer-Based (Figure~\ref{fig:eval}C).
A paired Wilcoxon signed-rank test confirms that the warm-start
advantage over the best static baseline (Throughput) is statistically
significant ($p{=}0.001$).
The warm-start prior biases toward BOLA, which exploits the
sustained medium-bandwidth corridors common in Norwegian urban
transit, mirroring the Puffer~\cite{yan2020puffer} finding that
small in-situ data can decisively shift policy selection without
full retraining.

%%
%% 4. CONCLUSION & FUTURE WORK
%%
\section{Conclusion and Future Work}
\label{sec:conclusion}

eBandit demonstrates that relocating ABR policy selection into the
Linux kernel via eBPF is both practical and impactful, composing
complementary heuristics without manual threshold engineering.
Future work includes a \emph{Contextual Bandit}~\cite{alt2019cba}
upgrade, Mahimahi evaluation, and porting shock detection into eBPF.

\balance
{\small
\bibliographystyle{ACM-Reference-Format}

\begin{thebibliography}{12}

\bibitem{abbasloo2020orca}
S.~Abbasloo, C.-Y. Yen, and H.~J. Chao.
\newblock Classic meets modern: {A} pragmatic learning-based congestion
  control for the {Internet}.
\newblock In \emph{Proc. ACM SIGCOMM}, pp. 632--647, 2020.

\bibitem{alt2019cba}
B.~Alt, T.~Ballard, R.~Steinmetz, H.~Koeppl, and A.~Rizk.
\newblock {CBA}: Contextual quality adaptation for adaptive bitrate video
  streaming.
\newblock In \emph{Proc. IEEE INFOCOM}, pp. 1--9, 2019.


\bibitem{huang2014bba}
T.-Y. Huang, R.~Johari, N.~McKeown, M.~Trunnell, and M.~Watson.
\newblock A buffer-based approach to rate adaptation: Evidence from a large
  video streaming service.
\newblock In \emph{Proc. ACM SIGCOMM}, pp. 187--198, 2014.

\bibitem{mao2017pensieve}
H.~Mao, R.~Netravali, and M.~Alizadeh.
\newblock Neural adaptive video streaming with {Pensieve}.
\newblock In \emph{Proc. ACM SIGCOMM}, pp. 197--210, 2017.

\bibitem{miano2023sketching}
S.~Miano, A.~Sanaee, F.~Risso, and G.~Antichi.
\newblock Fast in-kernel traffic sketching in {eBPF}.
\newblock \emph{ACM SIGCOMM Comput. Commun. Rev.}, 53(1):2--9, 2023.

\bibitem{riiser2013}
H.~Riiser, P.~Vigmostad, C.~Griwodz, and P.~Halvorsen.
\newblock Commute path bandwidth traces from 3{G} networks: Analysis and
  applications.
\newblock In \emph{Proc. ACM MMSys}, pp. 114--118, 2013.

\bibitem{spiteri2016bola}
K.~Spiteri, R.~Urgaonkar, and R.~K. Sitaraman.
\newblock {BOLA}: Near-optimal bitrate adaptation for online videos.
\newblock In \emph{Proc. IEEE INFOCOM}, pp. 1--9, 2016.

\bibitem{vieira2020ebpf}
M.~A.~M. Vieira, M.~S. Castanho, R.~D.~G. Pac\'{i}fico,
  E.~R.~S. Santos, E.~P.~M. C\^{a}mara~J\'{u}nior, and L.~F.~M. Vieira.
\newblock Fast packet processing with {eBPF} and {XDP}: Concepts, code,
  challenges, and applications.
\newblock \emph{ACM Comput. Surv.}, 53(1):16:1--16:36, 2020.

\bibitem{yan2020puffer}
F.~Y. Yan, H.~Ayers, C.~Zhu, S.~Fouladi, J.~Hong, K.~Zhang, P.~Levis,
  and K.~Winstein.
\newblock Learning in situ: A randomized experiment in video streaming.
\newblock In \emph{Proc. USENIX NSDI}, pp. 495--510, 2020.

\bibitem{yin2015mpc}
X.~Yin, A.~Jindal, V.~Sekar, and B.~Sinopoli.
\newblock A control-theoretic approach for dynamic adaptive video streaming
  over {HTTP}.
\newblock In \emph{Proc. ACM SIGCOMM}, pp. 325--338, 2015.

\bibitem{zhou2023electrode}
Y.~Zhou, Z.~Wang, S.~Dharanipragada, and M.~Yu.
\newblock Electrode: Accelerating distributed protocols with {eBPF}.
\newblock In \emph{Proc. USENIX NSDI}, pp. 1597--1613, 2023.

\end{thebibliography}

}

\end{document}